\documentclass[%
 reprint, 
superscriptaddress,
 amsmath,amssymb,
 aps,
hyperref
]{revtex4-2}
\usepackage{graphicx}
\usepackage{dcolumn}
\usepackage{bm}
\usepackage[colorlinks,citecolor=blue,urlcolor=blue,linkcolor=blue]{hyperref}
\begin{document}
\title{The Diffuse Supernova Neutrino Background \\ in the Standard and Double Collapse Models}
\author{Alexander Libanov}
\email[\bf{e-mail:  }]{libanov.am18@physics.msu.ru}
\author{Andrey Sharofeev}%
\email[\bf{e-mail:  }]{sharofeev@ms2.inr.ac.ru}
\affiliation{%
Faculty of Physics, Lomonosov State University, 1-2 Leninskiye Gory, 119991, Moscow, Russia
}%
\affiliation{Institute for Nuclear Research of the Russian Academy of Sciences, 60th October Anniversary Prospect 7a, Moscow 117312, Russia}
\date{\today}
\begin{abstract}
 The diffuse supernova neutrino background (DSNB) is a powerful future tool to constrain core-collapse explosion mechanisms without observation of a nearby event, and the corresponding signal has been calculated for a variety of collapse models. For Supernova (SN) 1987A, a peculiar double neutrino burst was detected, but models for the double collapse have never been studied in the DSNB context. Here, we fill this gap and compare the DSNB signal expected in the Standard Collapse (SC) and the Double Collapse (DC) models in various future detectors, including Hyper-Kamio\-kande, JUNO, DUNE and the Large Baksan Neutrino Telescope (LBNT). We calculate the spectra of diffuse neutrinos and antineutrinos in the DC model and determine the rate of registered events as a function of energy of the detected particle, taking into account detector parameters. For each detector, we estimate the corresponding uncertainties and the background and compare the signals expected for the SC and DC models. We conclude that the combination of DUNE and LBNT data will have the highest sensitivity to discriminate between the SC and DC models.
\end{abstract}
\maketitle
\section{Introduction}
\label{intro}
Core-Collapse Supernova (CCSN) is an event which is the final stage of a stellar evolution ($M > 8M_{\odot}$, where $M_{\odot}$ is the mass of the Sun, $M$ is the stellar mass of the progenitor star). At the beginning of the 20th century it was predicted that during extreme processes in the interior of massive stars before their explosion, radiation of neutrinos and antineutrinos should occur. In the last seconds of its life, the star emits a large number of neutrinos and antineutrinos $\sim 10^{58}$, releasing almost all of the star's energy $\sim 10^{53}$ erg \cite{Beacom_2010} (see also \cite{Horowitz_2001, Page_2006, LATTIMER_2007}). The spectrum of neutrinos from CCSNe makes it possible to get closer to understanding the final stage of stellar evolution \cite{JANKA_2007, Woosley_2005, Lunardini_2009}. However, the rarity of nearby events ($ \sim 3$ events per century \cite{doi:10.1146/annurev.nucl.010909.083331,2015ApJ...802...47A,2021NewA...8301498R}) makes the prospect of direct registration of neutrinos and antineutrinos from a single event uncertain. Direct registration of the diffuse neutrino flux would allow one not to wait for such a rare event, but to get knowledge about the stellar evolution, collecting the neutrino background from the entire Universe. To this day, Diffuse Supernovae Neutrino Background (DSNB) remains undetected experimentally while theoretical predictions suffer from large uncertainties in the CCSNe rate and explosion models, caused by the relative rarity of nearby supernovae and the presence of only one event, SN 1987A, detected in the neutrino channel.

The only registered SN neutrino detection occurred in 1987, and the observed neutrino light curve from SN 1987A had two peaks separated in time by $\sim 4.7$ hours \cite{Aglietta:1987it,Alekseev:1987ej,1987PhRvL..58.1490H,1987PhRvL..58.1494B}. The double neutrino signal the supernova SN 1987A can not be explained within the standard collapse scenario. Based on the available data, models have been developed to explain the cause of the unusual time distribution of neutrinos \cite{Berezinsky:1988qca}. Recent studies suggest that the number of SNe subjected to non-standard scenario is about $(0.1 - 1)\%$ of the total core collapses \cite{Postnov_2016}, which forces us to turn to alternative models, for instance, the mag\-neto-rotational mechanism model \cite{Bisnovatyi_Kogan_2018} or the rotational mechanism \cite{Imshennik_2004,Imshennik_2010,Ryazhskaya_2006}, which will be considered in this study.

Measurement of the DSNB signal will help to discriminate between various scenarios of core collapses \cite{Lunardini_2012}, \cite{M_ller_2018}. However, the double collapse model has not been considered in the context of the DSNB yet, and the present work fills this gap. We estimate below the capabilities of relevant future instruments such as Hyper-Kamiokande (HK), JUNO, DUNE and the Large Baksan Neutrino Telescope (LBNT) \cite{kamiokandecollaboration2021diffuse, Novoseltsev_2020, Migenda_2017, junocollaboration2021juno, dunecollaboration2021design}.
Based on the analysis of the expected parameters of the detectors, it will be concluded that it is possible to distinguish the models in the course of future experiments, as well as to choose the most suitable detector for this purpose.

\section{The Rate of Core-Collapse Supernovae}
\label{Core Collappse Supernova Rate}

\begin{figure}
\includegraphics[scale=0.51]{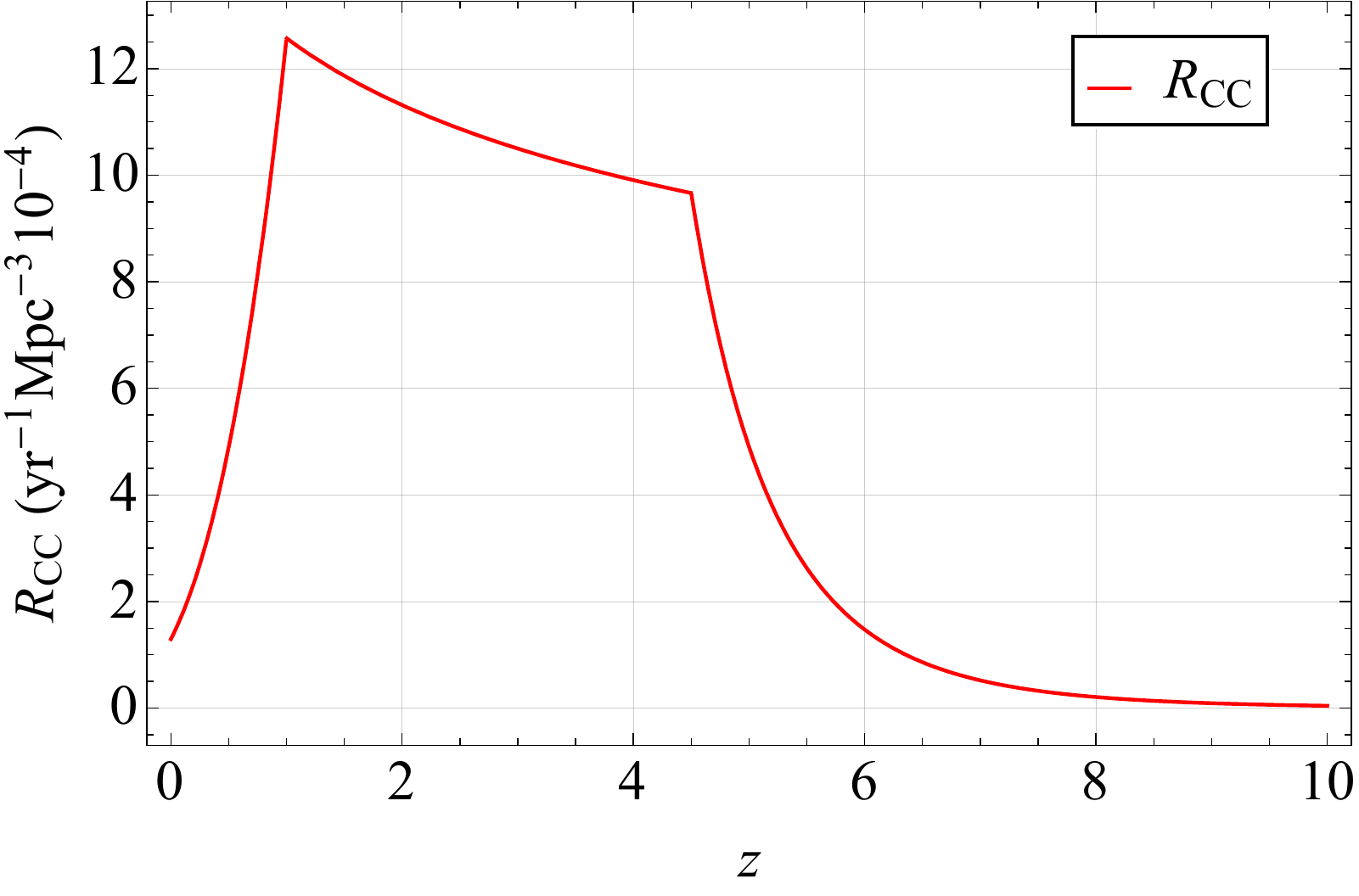}
\caption{Core collapse supernovae rate \cite{Lunardini_2016} with normalization given in \cite{Ando_2004} (red line),  $z$ is the redshift.}
\label{RccGraph}       
\end{figure}

The CCSNe rate is known up to large uncertainties, related in particular to the overall normalization. 
Various approximations \cite{Anandagoda_2020,Y_ksel_2008,Madau_2017} are commonly used. 
We focus on the approximation proposed in \cite{Lunardini_2016} with the normalization obtained as a result of observations of the nearest supernovae stars \cite{Hopkins_2006,Ando_2004,Strolger_2015},  
\begin{equation}
\begin{gathered}
    \label{eq1.1}
    R_{CC}(z) = R_{-4} \times \frac{10^{-4}}{\mbox{yr}\cdot \mbox{Mpc}^{3}} ((1+z)^{\beta}\Theta(z)\Theta(1-z)+ \\
    +2^{\beta-\alpha}(1+z)^{\alpha}\Theta(z-1)\Theta(4.5-z)+ \\
+2^{\beta-\alpha}5.5^{\alpha-\gamma}(1+z)^{\gamma}\Theta(z-4.5)),
\end{gathered}
\end{equation}
where $\alpha = -0.26$, $\beta = 3.28$, $\gamma = -7.8$, $z$ is the redshift, $\Theta(z)$ is the Heaviside Theta function, $R_{-4} \approx 1.2$ is the normalization. The CCSNe rate (\ref{eq1.1}) is shown in Fig. \ref{RccGraph}.
\section{Collapse models}
\label{Collapse Models}
Here, we concentrate on the general features of the neutrino and antineutrino emission and not on details which vary depending on particular details of the supernova models. We will compare the Standard Collapse and the Double collapse models. Their distinctive features with respect to the neutrino and antineutrino spectra are discussed in this section.    
\subsection{The Standard Collapse}
\label{Standard Collapse}
For the purpose of this work, we do not distinguish any variations of models of stellar collapse and subsequent supernova explosion which predict a single neutrino signal per event; we call them ''Standard Collapse'' (SC) models. Detailed descriptions of these models are available e.g. in Refs. \cite{Beacom_2010}, \cite{HARTMANN1997137}. Implications of variations and uncertainties of these models to DSNB have been widely discussed in \cite{Mirizzi_2005}.
The integrated spectrum of neutrinos and antineutrinos in the Standard Collapse is in good agreement with the thermal spectrum and is given by the following formula \cite{Beacom_2010}, \cite{Raffelt:1996wa}, \cite{Kotake_2006},
\begin{equation}
    \label{eq2.1}
    f^{SC}_{\nu_{i}}(E_{\nu_{i}},T_{\nu_{i}}) = \frac{E^{tot}_{\nu}}{6}\frac{120}{7\pi^{4}}\frac{E^{2}_{\nu_{i}}}{T^{4}_{\nu_{i}}}(e^{E_{\nu_{i}}/T_{\nu_{i}}}+1)^{-1},
\end{equation}
where $E^{tot}_{\nu} \approx 3 \times 10^{53} \mbox{erg}$ is the total energy carried by neutrinos and antineutrinos of all flavors. This energy is equal to the difference in gravitational energy during the collapse of a star \cite{Horowitz_2001, Page_2006, LATTIMER_2007}, $E_{\nu_{i}}$ is the neutrino and antineutrino energy near the stellar surface,  $T_{\nu_{i}}$ is the temperature of the neutrinosphere \cite{Horiuchi_2009} (further, typical neutrinosphere temperatures are $T_{\nu_{e}}= 5 \mbox{ MeV, }T_{\bar{\nu}_{e}} = 6\mbox{ MeV,}$ $T_{\nu_{\mu}}=T_{\nu_{\tau}}=T_{\bar{\nu}_{\mu}}=T_{\bar{\nu}_{\tau}}=7\mbox{ MeV}$ \cite{Mathews_2014}), $i$ denotes the flavor. It is assumed that neutrinos and antineutrinos carry away approximately the same fraction of gravitational energy in this case \cite{Beacom_2010}.

\subsection{The Double Collapse}
\label{The Double Collapse}
We use the most elaborated DC model described in detail in Refs. \cite{Imshennik_2004}, \cite{Imshennik_2010}, \cite{Ryazhskaya_2006}. In this model, fast rotation of neutron-star progenitor results in core deformation into disk which splits into two small neutron stars. 

The first neutrino burst is associated with the explosion of the smaller neutron star which becomes unstable after accretion of matter on the bigger one and occurs within rotational mechanism  \cite{Imshennik_2004,Imshennik_2010,Ryazhskaya_2006}. The second neutrino burst is associated with the collapse of
the formed massive neutron star. 

Hence, the total contribution of a supernova to the DSNB in this model is the sum of these two signals. 

\subsubsection{Neutrinos from The First Burst}
\label{Neutrinos from The First Burst}
When a matter is neutronized during the stellar collapse, 
\[
e^{-}+p \to n + \nu_{e},
\]
only electron neutrinos are born.

Neutrino and antineutrino of other flavors are born during the rescattering of electron neutrinos in the opaque area. Hence, the spectra of electron neutrinos from the transparent area and neutrinos and antineutrinos of other flavors from the opaque area are different.    

Electron neutrinos from the transparent area spectrum and electron antineutrinos (and other flavors and types except electron neutrinos) from the opaque area spectrum can be found from the approximation of the plots presented in  \cite{Imshennik_2004}, \cite{Imshennik_2010}. The expression for the $\nu_{e}$ spectrum in units $\mbox{MeV}^{-1}$ from the transparent region is

\begin{equation}
f_{rot,1}(E _{\nu _{e}}) \approx
\chi_1 \left( \frac{E_{\nu _{e}}}{\mbox{MeV}} \right)^5\Bigg(1+\exp\Bigg({\frac{E_{\nu
_{e}}}{\omega_1}}\Bigg)\Bigg)^{-1},
\label{Eq/Pg3/2:physcourse}
\end{equation}
and the $\nu_{i},\bar{\nu}_{i}$ spectra from the opaque region are

\begin{equation}
f_{rot,2}(E_{\nu_{i},\bar{\nu}_{i}}) \approx \chi_2 \left(\frac{E_{\nu_{i},\bar{\nu}_{i}}}{\mbox{MeV}}\right)^{5} \Bigg( 1 + \exp\Bigg( \frac{E_{\nu_{i},\bar{\nu}_{i}}}{\omega_2} \Bigg) \Bigg)^{-1},
\label{eq2.2}
\end{equation}
where $E_{\nu_{e}}$, $E_{\nu_{i},\bar{\nu}_i}$ are in MeV; $\omega_1 = 6.32 \mbox{ MeV}$, $\omega_2 = 6.28 \mbox{ MeV}$; $\chi_1 = 2.33 \times 10^{51}$, $\chi_2 = 1.29 \times 10^{50}$ are dimensionless.

\subsubsection{Neutrinos from The Second Burst}
\label{Neutrinos from The Second Burst}
The second neutrino burst presumably occurs \cite{Ryazhskaya_2006} within the SC described in section \ref{Standard Collapse}. Note that during the neutronization of the matter only electron neutrinos are born and neutrino and antineutrino of other flavors are born during the electron neutrinos rescattering on the opaque area.

\subsubsection{The Double Collapse Spectrum}
\label{The Double Collapse Spectrum}
The total energy of the two bursts is
\begin{equation}
\alpha (E_{rot,1} + 6E_{rot,2})+\beta (6E_{\nu ,\bar\nu } -
(E_{rot,1} + 6E_{rot,2})),
\label{eq2.4}
\end{equation}
where $E_{rot,1} = 3.14\times10^{52} \mbox{erg}$ is the $\nu_{e}$ energy from transparent region, $E_{rot,2} = 1.7\times10^{52} \mbox{erg}$ is that carried away by the only one neutrino and antineutrino flavor after the rescattering \cite{Imshennik_2010}, and $\alpha $ and $\beta $ are the undefined coefficients.

The ratio of these coefficients is equal to the ratio of the total energies in the first and
second signals,
\begin{equation}
\frac{\alpha }{\beta } = \frac{E_{rot,1}+6E_{rot,2}}{6E_{\nu ,\bar\nu }
- (E_{rot,1} + 6E_{rot,2})}.
\label{eq2.5}
\end{equation}
The total energy does not depend on the choice of the model, so we equate
(\ref{eq2.4}) to $ 6E_{\nu,\bar\nu}$. Solving the system  (\ref{eq2.4}) and (\ref{eq2.5}) with respect to $\alpha$ and
$\beta$, we find:   
 \begin{equation}
 \begin{cases}
\alpha \approx 0.17
\\
\beta \approx 1.05.
\end{cases}
\label{eq2.6}
\end{equation}

The electron neutrino spectrum in the case of the DC becomes
\begin{equation}
f_{\nu _{e}}^{DC}(E_{\nu _{e}}) = \alpha(f_{rot,1}(E_{\nu _{e}}) +
f_{rot,2}(E_{\nu _{e}})) + \beta
f^{SC}(E_{\nu _{e}},T_{\nu_{e}}).
\label{eq2.7}
\end{equation}
And the spectrum of neutrinos and antineutrinos of other flavors (including
electron antineutrinos) in the case of the DC are given by following formula.
\begin{equation}
f_{\nu _{i},\bar{\nu}_{i} }^{DC}(E_{\nu _{i}, \bar{\nu }_{i}}) =
\alpha f_{rot,2}(E_{\nu _{i}, \bar{\nu }_{i}}) + \beta f^{SC}(E_{\nu _{i},
\bar{\nu }_{i}},T_{\nu _{i},
\bar{\nu }_{i}}),
\label{eq2.8}
\end{equation}
where $i$ denotes the flavor.

\section{The Mikheyev-Smirnov-Wolfenstein Effect}
\label{MSW-effect}
The Mikheyev-Smirnov-Wolfenstein (MSW) effect is the effect of 
transformation of one neutrino species (flavor) into another one in a medium with varying density \cite{smirnov2003msw}. 
During this process, neutrinos (hereinafter, the normal hierarchy of neutrino masses is assumed) pass from one type to another according to the following formulas for the SC \cite{Tabrizi_2021,Lunardini_2012,Porto_Silva_2021, Lu_2016}:
\begin{align}
        f_{\nu_{e}}^{SC,MSW}  =  f_{\nu_x}^{SC},
        \label{MSW1SC}
        \\
        f_{\bar{\nu}_{e}}^{SC,MSW}  =  c_{12}^{2} f_{\bar{\nu}_{e}}^{SC} + s_{12}^{2} f_{\nu_{x}}^{SC},
        \label{MSW2SC}
        \\
        f_{\nu_{x}}^{SC,MSW}  =  \frac{1}{4}[f_{\nu_{e}}^{SC} + s_{12}^{2} f_{\bar{\nu}_{e}}^{SC} + (2 + c_{12}^{2}) f_{\nu_{x}}^{SC}],
    \label{MSW3SC}
\end{align}
where $c_{12}^{2} = 1 - s_{12}^{2}$ and $s_{12}^{2} = 0.310$ are the mixing angles, $f_{\nu}$ is the neutrino flux without the MSW-effect, $\nu_{x}$ is muon or tau neutrino. The neutrino and antineutrino spectra for the DC with the MSW effect are given by the following formulas.

\begin{align}
        f_{\nu_{e}}^{DC,MSW}  =  \alpha(f_{rot,1}(E_{\nu _{e}}) +
f_{rot,2}(E_{\nu _{e}})) + \beta f_{\nu_{e}}^{SC,MSW} \label{MSW1DC}, \\
        f_{\bar{\nu}_{e}}^{DC,MSW}  = \alpha f_{rot,2}(E_{\bar{\nu} _{e}}) + \beta  f_{\bar{\nu}_{e}}^{SC,MSW},
        \label{MSW2DC}
        \\
        f_{\nu_{x}}^{DC,MSW}  =   \alpha f_{rot,2}(E_{\nu_{x}})+\beta f_{\nu_{x}}^{SC,MSW}.
    \label{MSW3DC}
\end{align}

It should be noted, that the MSW-effect is already taken into account in the first terms of the formulas (\ref{MSW1DC}) -- (\ref{MSW3DC}), since this is the neutrino spectrum after interaction with the stellar matter \cite{Imshennik_2010}.

The same formulas work for antineutrinos if we take into account that there is no mixing between neutrinos and antineutrinos.

\section{The Neutrino Oscillations}
\label{Neutrino Oscillations}
After the neutrinos and antineutrinos escape from the collapsed star, they propagate in vacuum. It is at this stage that the effect of vacuum oscillations of neutrinos and antineutrinos becomes essential. Oscillations of electron antineutrinos are  accounted following \cite{Palladino:2020jol} (see also \cite{Lunardini_2012})
\begin{equation}
    \label{OscSC}
    f^{osc,SC}_{\nu_{e},\bar{\nu}_{e}} = 0.548f^{SC,MSW}_{\nu_{e},\bar{\nu}_{e}}+0.185f^{SC,MSW}_{\nu_{\mu},\bar{\nu}_{\mu}}+0.267f^{SC,MSW}_{\nu_{\tau},\bar{\nu}_{\tau}},
\end{equation}
where $f_{\nu_{i}, \bar{\nu_{i}}}^{SC,MSW}$ are fluxes of neutrino and antineutrino with the MSW-effect for the SC and
\begin{equation}
    \label{OscDC}
    f^{osc,DC}_{\nu_{e},\bar{\nu}_{e}} = 0.548f^{DC,MSW}_{\nu_{e},\bar{\nu}_{e}}+0.185f^{DC,MSW}_{\nu_{\mu},\bar{\nu}_{\mu}}+0.267f^{DC,MSW}_{\nu_{\tau},\bar{\nu}_{\tau}}
\end{equation}
for the DC, where $f_{\nu_{i}, \bar{\nu_{i}}}^{DC,MSW}$ are fluxes of neutrino and antineutrino with the MSW-effect for the DC.

\section{Total Neutrino and Antineutrino Fluxes on the Earth}
\label{Total Fluxes on the Earth}
\begin{figure}
\includegraphics[scale=0.51]{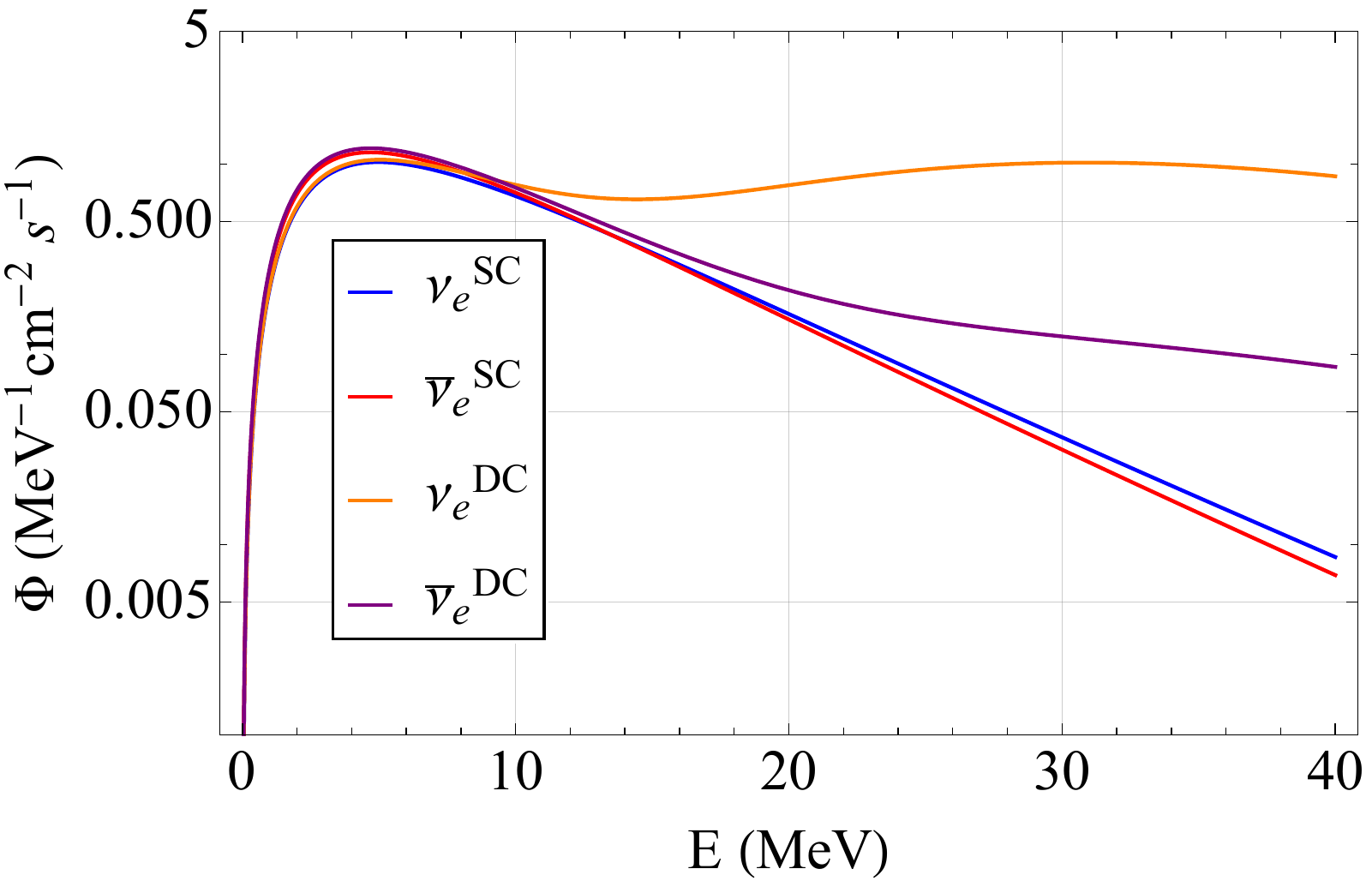}
\caption{The electron neutrino and electron antineutrino fluxes at the Earth in various models. The blue line is the flux of electron neutrinos in case of the SC, the red line is the flux of electron antineutrinos in case of the SC, the orange line is the flux of electron neutrinos in case of the DC and the purple line is the flux of electron antineutrinos in case of the DC.}
\label{Fluxes}      
\end{figure}

The electron neutrino and antineutrino fluxes are the most interesting in context of this study and, hence, observed on Earth fluxes are given by the following the formula \cite{Ando_2004_2},

\begin{equation}
\Phi^{\nu_{e},\bar{\nu}_{e}}_{m} (E_{0},T) = \frac{c}{H_{0}}\int
\limits_{0}^{z_{max}}\frac{R_{CC}f_{\nu_{e},\bar{\nu}_{e}}^{osc,m}(E_{0}(1+z),T)}{\sqrt{\Omega_{M}(1+z)^{3}+\Omega_{\Lambda}}} dz,
\label{Flux}
\end{equation}
where $c = 3 \cdot 10^{10}$ $\mbox{cm } \, \mbox{s}^{-1}$ is the speed of light in a vacuum,
$H_{0} = 2.19 \cdot 10^{-18}$ s$^{-1}$ is the present Hubble constant, $\Omega_{M} = 0.315$ is the energy density of non-relativistic matter,
$\Omega_{\Lambda} = 0.685$ is the dark energy density \cite{Planck_2020}, index $m$ refers to the collapse model (SC or DC), $E_{0}$ is energy of neutrino and antineutrino at the Earth. The limit of integration $z_{max} = 5$ was chosen from the fact that SNe rate is estimated with a reasonable accuracy up to these redshifts, although the result almost does not depend on the high redshifts cut-off. Spectra of electron neutrinos and antineutrinos for various models are shown in Fig. \ref{Fluxes}.

\section{The DSNB Detectors}
\label{DSNB detectors}

To register DSNB, it is planned to use two main types of detectors: liquid scintillators and Cherenkov detectors. The energies of neutrinos and antineutrinos have characteristic values of the order of $10$ MeV. Due to the large errors associated with the CCSNe rate and the specifics of the detectors (we remind that all the detectors we discuss are at their project stage currently ), we take an approximate estimate of the error in determining the number of events $\pm 50\%$. This conservative estimate is in good agreement with those presented in Refs. \cite{Tabrizi_2021} and \cite{M_ller_2018}. Detectors are characterized by several general parameters: the registration channel, filling, fudicial volume, the number of targets, the registration energy range and finally the efficiency.

Let us dwell in more detail on the cross-section of the registration channel. The main filling in detectors is either water (or water with gadolinium), linear  alkylbenzene for liquid-scintillator detectors, or liquid argon. The main registration channel is Inverse Beta-Decay (IBD) in the case of water (or linear  alkylbenzene),

\begin{equation}
\label{IBD}
  \bar{\nu}_{e} + p \to n + e^{+}.
\end{equation}

\begin{figure}
\includegraphics[scale=0.51]{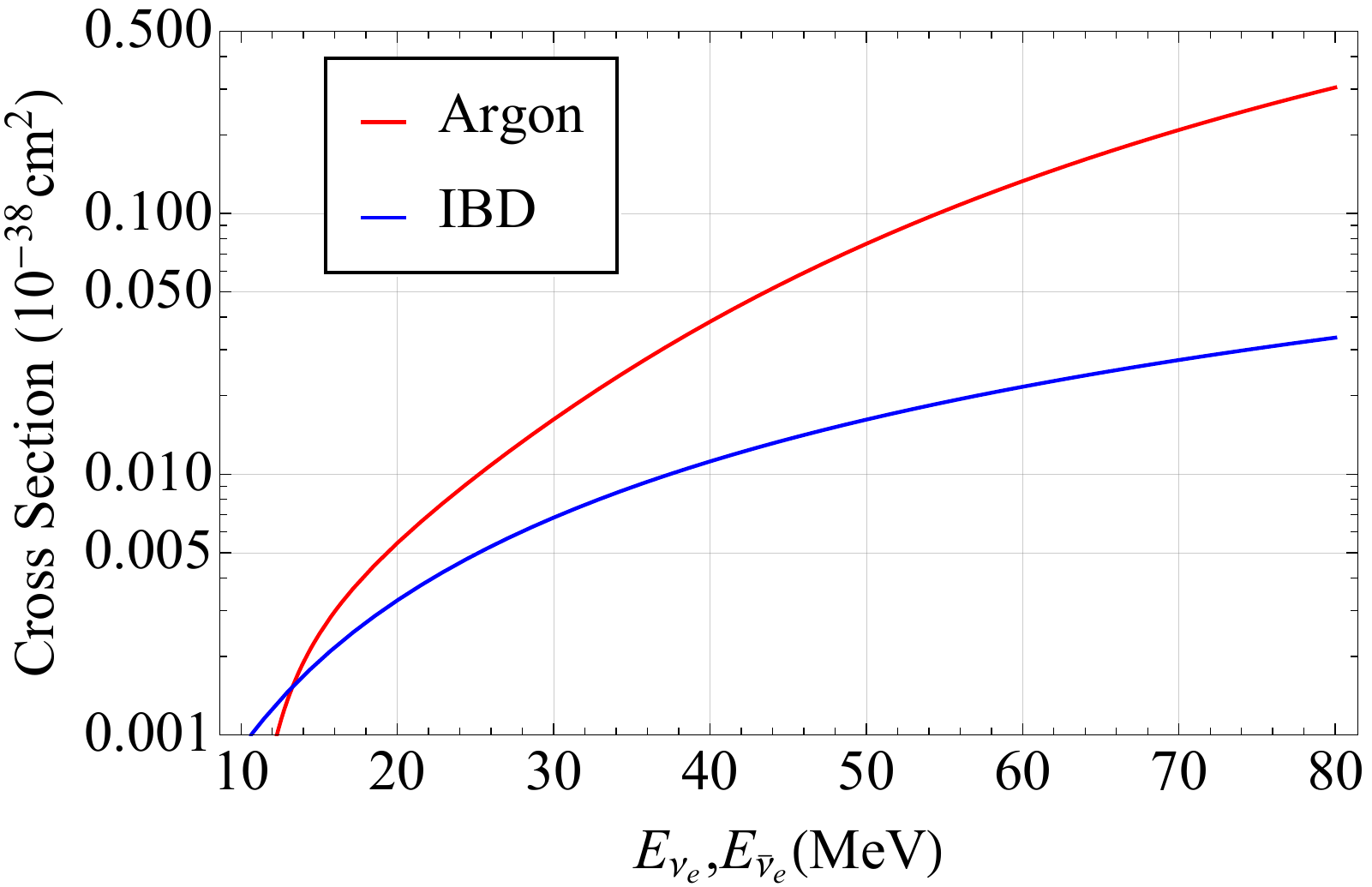}
\caption{IBD reaction (\ref{IBD}) and argon reaction (\ref{ArgonReaction}) cross-sections. The blue line is the IBD-cross-section and the red line is the argon reaction cross-section.}
\label{fig1}       
\end{figure}

\begin{table*}
\caption{Main parameters of the detectors considered \cite{Tabrizi_2021}, \cite{Bayarto}.}
\label{detectors}
\begin{ruledtabular}
\begin{tabular*}{\textwidth}{@{\extracolsep{\fill}}lllllll@{}}
Experiment & \multicolumn{1}{c}{Registration Channel} & \multicolumn{1}{c}{Filling} & \multicolumn{1}{c}{Fiducial Mass (kt)} & \multicolumn{1}{c}{Targets} & Energy Range (MeV) & Efficiency($\%$) \\
\hline
  HK(Gd) & IBD & Water + Gd & 374 & $2.5\times 10^{34}$  & 12-24 & 67
        \\
        JUNO  & IBD & $\mbox{C}_{6}\mbox{H}_{5}\mbox{C}_{12}\mbox{H}_{25}$ & 17 & $1.21\times 10^{33}$ & 10-22 & 50
        \\
        DUNE & $\nu_{e}+Ar$ & Ar & 40 & $6.02\times 10^{32}$ & 19-32 & 86
        \\ 
        LBNT & IBD & $\mbox{C}_{6}\mbox{H}_{5}\mbox{C}_{12}\mbox{H}_{25}$  & 10 & $7.3\times 10^{32}$  & ~0-100 & 95-100
        \\
\end{tabular*}
\end{ruledtabular}
\end{table*}
Some detectors such as DUNE and ICARUS use argon as a filling. The reaction with argon, which detects electron neutrinos, is given below

\begin{equation}
\label{ArgonReaction}
\nu _{e} + ^{40}Ar \to e^{-} + ^{40}K^{*}. 
\end{equation}

Cross-sections of these reactions as functions of electron (anti)neutrino energies are shown in Fig. \ref{fig1}  \cite{Formaggio_2012}, \cite{Tabrizi_2021}, \cite{abi2020deep}. 

Thus, we can select the most promising detectors and evaluate their future performance over 20 years. In this paper, four detectors will be considered: HK, JUNO, DUNE and the LBNT. Another promising experiment to measure DSNB is planned to be installed in the Jing Ping Laboratory but we could not find sufficient information about this future detector to calculate the number of events in it.
To calculate the number of events, we need expressions for the neutrino and antineutrino fluxes observed at the Earth (\ref{Flux}), calculated with the account of the MSW-effect (\ref{MSW1SC}) - (\ref{MSW3DC}) and oscillations (\ref{OscSC}), (\ref{OscDC}) and the parameters of the detectors presented in the Table \ref{detectors}.

It should be noted that electron antineutrinos (like any other flavor of
neutrinos and antineutrinos) are observed indirectly due to difficulty of their direct registration (for example, according to the positrons they generate and according to Cherenkov radiation).
\begin{equation}
N = \varepsilon N_{t}t \int \Phi(E,T) \sigma
_{i}(E)dE,
\label{EventRateFormula}
\end{equation}
where $\varepsilon $ is the efficiency ($\%$), $N_{t}$ is the number of targets, $\Phi (E)$
is the flux electron (positron) neutrino at the Earth ($\mbox{eV}^{-1} \, \mbox{cm}^{-2} \, \mbox{s}^{-1}$), $\sigma_{i}(E)$ is the cross-section $(\mbox{cm}^{2})$, the index $i$ refers to a registration channel, $t=20 \, \mbox{ yrs}$ is the total registration time \cite{M_ller_2018}, \cite{Tabrizi_2021}.  
\section{The Total Event Rate}
\label{Total Event Rate}
\subsection{Hyper-Kamiokande}
\label{HK}
HK is a water detector with the addition of water-soluble gadolinium (Gd) sulfate which is under construction in Japan as a successor to the Super-Kamiokande. At present, it is planned that the detector will contain two cylindrical tanks with the fiducial mass of 187 kt each. The main reaction for registering electron antineutrinos is IBD. It has been calculated that adding $GdCl_{3}$ to water will reduce the muon background by 5 times \cite{M_ller_2018}, \cite{protocollaboration2018hyperkamiokande}. There are several backgrounds that significantly affect DSNB detection. These backgrounds arise as a result of the interaction of cosmic rays with the atmosphere and, as a consequence, the produced muon neutrinos and antineutrinos (charged current, CC), the emission of $\gamma$-rays that are born after the removal of excitations (which appear as a result of quasi-elastic interactions of cosmic rays with the atmosphere) of charged currents (NC), as well as the spallation of $^{9}$Li atoms. These backgrounds are expected to be non-removable \cite{Huang:2015hro}, \cite{Tabrizi_2021}.

\begin{figure}
\includegraphics[scale=0.51]{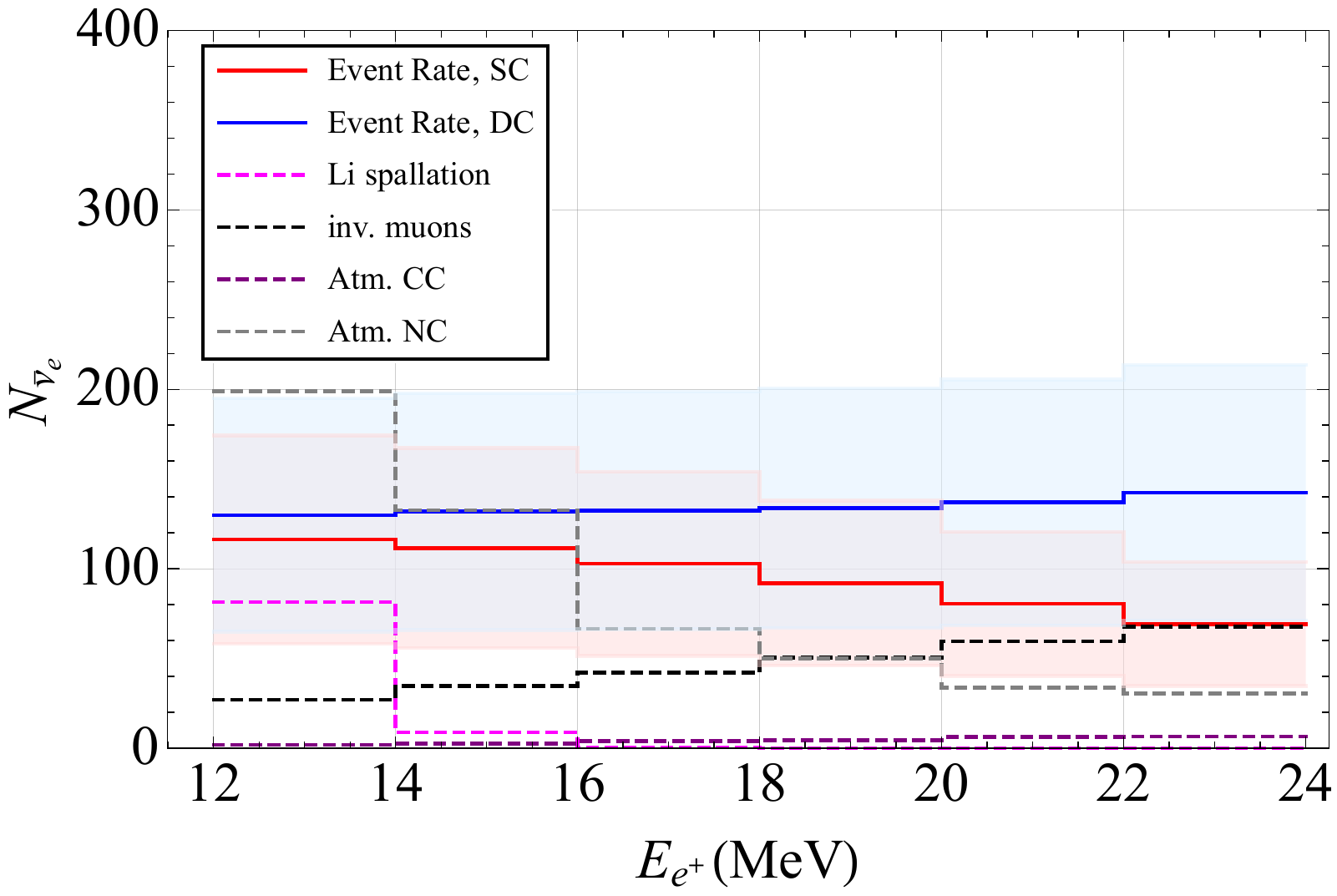}
\caption{Estimated number of events in 20 years for HK as a function of positron energy. Red and blue lines are SC and DC event rates, respectively. Magenta, black, purple and gray dashed lines are Li spallation, invisible muons, atmospheric CC and atmospheric NC backgrounds, respectively.}
\label{HKgraph}       
\end{figure}

Using the formula (\ref{EventRateFormula}) and taking into account the backgrounds, we can obtain a graph of the dependence of the number of events on the energies of positrons that are actually registered in this detector. This dependence with backgrounds is shown in Fig. \ref{HKgraph}.
\subsection{JUNO}
\label{JUNO}
Jiangmen Underground Neutrino Observatory (JUNO) is a 20 kt liquid-scintillator detector planned to be built in China \cite{An_2016}. The detector contains the central tank filled with  linear alkylbenzene. The central detector is immersed in the water Cherenkov detector surrounded by a muon tracker to reduce the muon background \cite{M_ller_2018}. The efficiency of this detector also strongly depends on the background of fast neutrons, but, nevertheless, it can be greatly reduced by reducing the volume to 17 kt \cite{An_2016}, since fast neutrons generated by cosmic muons should decay directly near the detector. Similar to HK, this detector is susceptible to background from CC and NC. 

\begin{figure}
\includegraphics[scale=0.51]{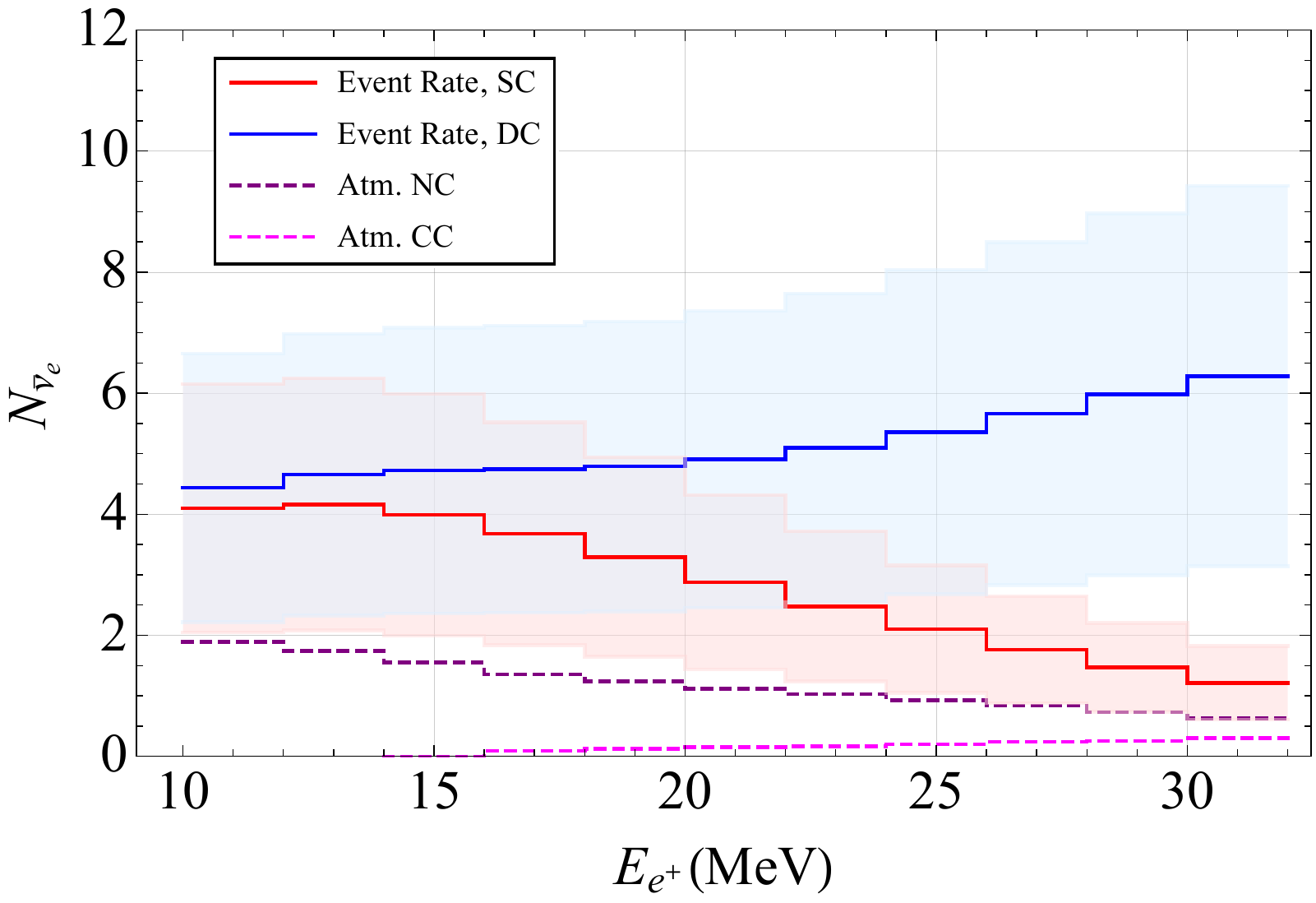}
\caption{Estimated number of events in 20 years for JUNO as a function of positron energy. Red and blue lines are SC and DC event rates, respectively. Purple and magenta dashed lines are atmospheric NC and CC backgrounds, respectively.}
\label{JUNOgraph}      
\end{figure}

The event rate calculated with (\ref{EventRateFormula}) and the background, are shown in Fig. \ref{JUNOgraph}.

\subsection{DUNE}
\label{DUNE}
The Deep Underground Neutrino Experiment (DUNE) is a detector based on 40 kt of liquid argon  to be built in South Dakota \cite{abi2020deep}, \cite{acciarri2016longbaseline}. Nowadays it is supposed to separate these
40 kilotons for 4 chambers of 10 kilotons. The main registration channel is,
\[
\nu _{e} + ^{40}Ar \to e^{-} + ^{40}K^{*}.
\]
It is planned that DUNE will have a trigger efficiency of about $90\%$, and a reconstruction efficiency $96\%$ hence $\varepsilon$ is $86\%$ \cite{M_ller_2018}.
\begin{figure}
\includegraphics[scale=0.51]{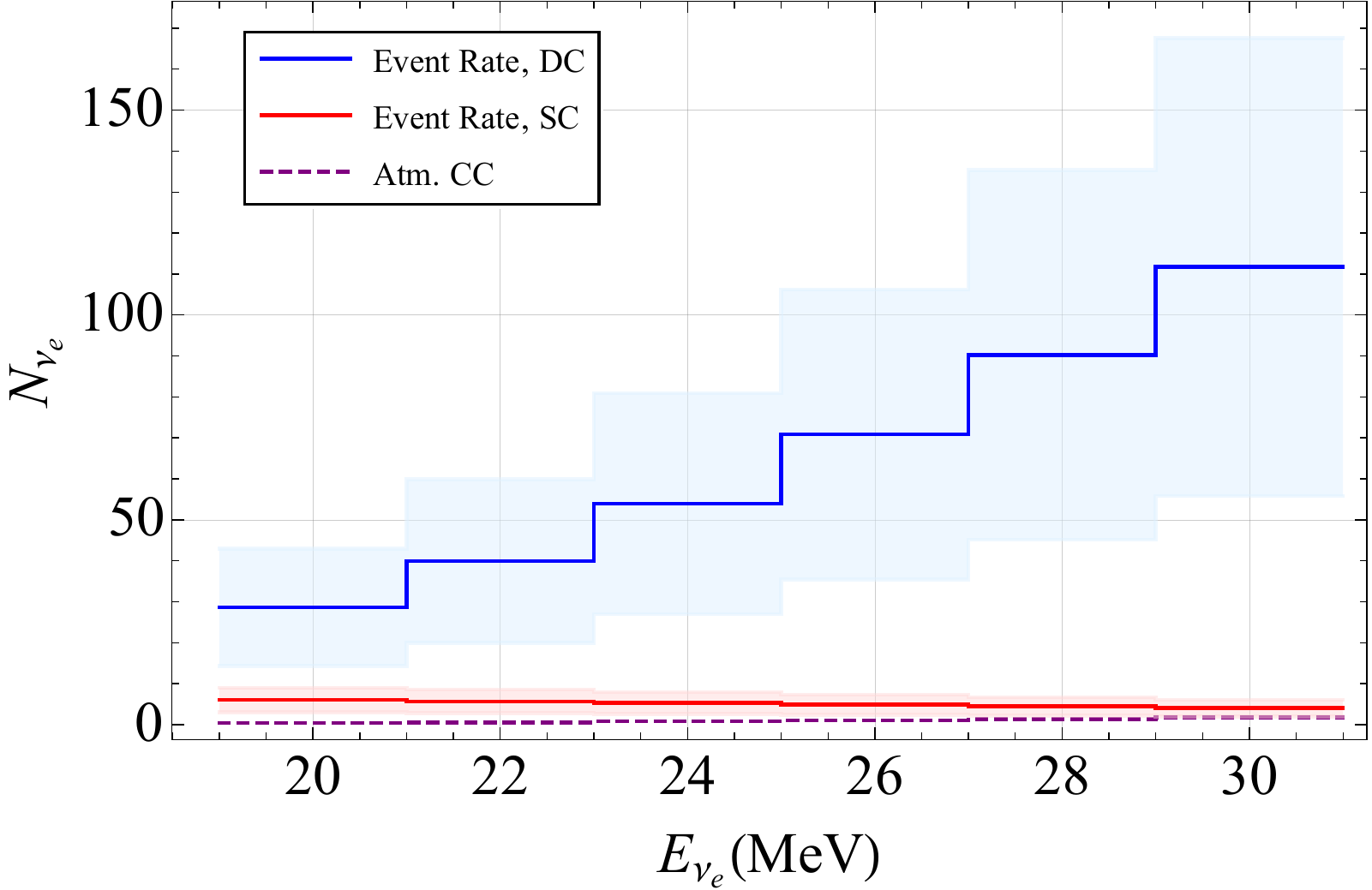}
\caption{Estimated number of events in 20 years for DUNE as a function of neutrino energy. Red and blue lines are SC and DC event rates, respectively. The dashed purple line is the atmospheric CC background.}
\label{DUNEDC}     
\end{figure}

\begin{figure}
\includegraphics[scale=0.51]{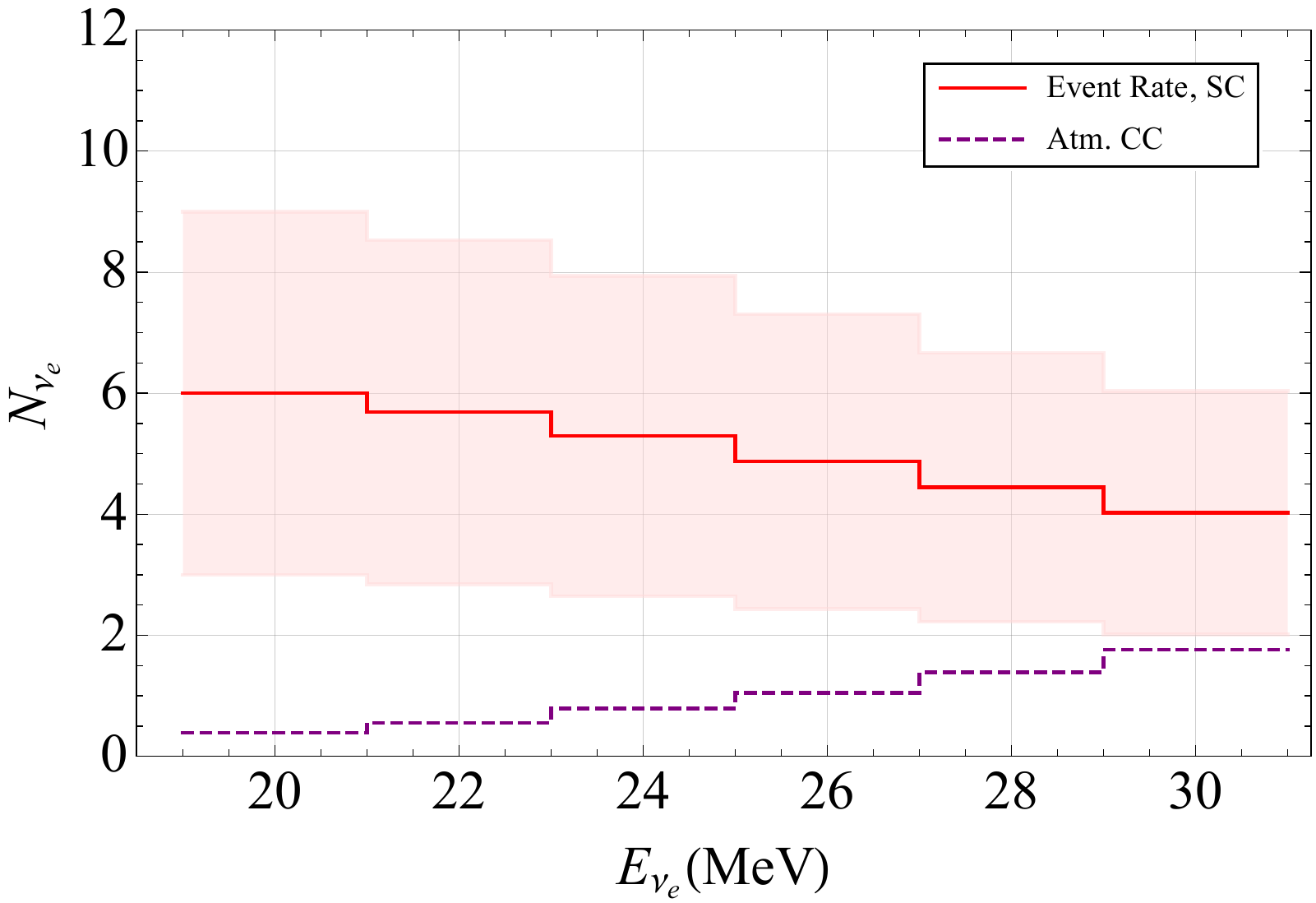}
\caption{Zoom of Fig. \ref{DUNEDC} for the SC curve.}
\label{DUNESM}    
\end{figure}
The structure and physics of the backgrounds for the DUNE are still 
under discussion,and thus we will use the assumption that the future DUNE and the current ICARUS \cite{M_ller_2018}, \cite{Cocco_2004} are 
similar, since both detectors use the same method. This detector is strongly influenced by backgrounds from solar neutrinos. Another irreducible background is the atmospheric neutrinos background \cite{Zhu_2019}.

The number of events as a function of electron neutrino energies is obtained in the similar way to HK and JUNO, taking into account backgrounds. The estimated event rates and the backgrounds are shown in Fig. \ref{DUNEDC}, \ref{DUNESM}.

\subsection{The Large Baksan Neutrino Telescope}
\label{The Large Baksan Neutrino Telescope}
\begin{figure}
\includegraphics[scale=0.51]{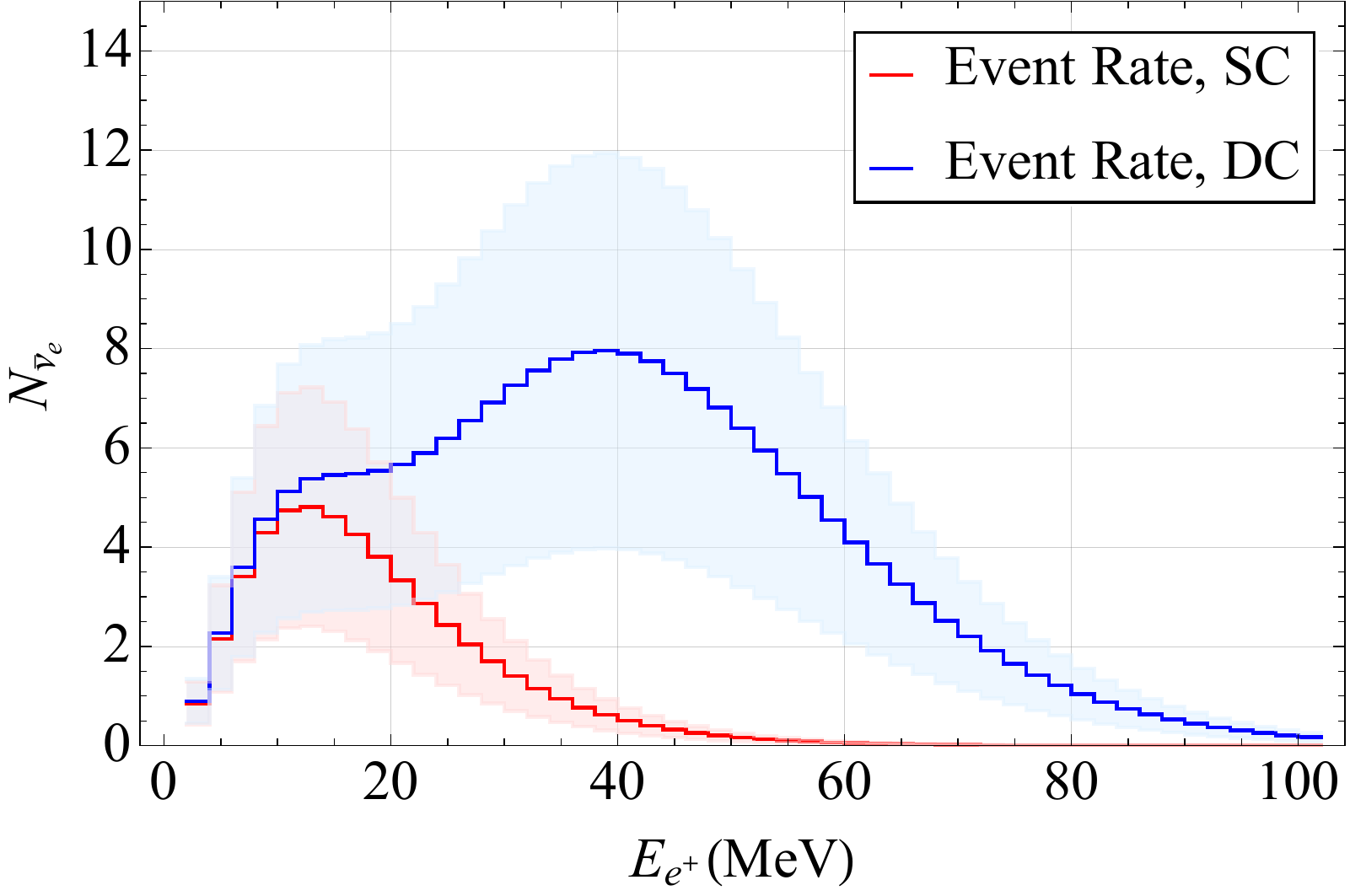}
\caption{Estimated number of events in 20 years for the LBNT as a function of positron energy. Red and blue lines are SC and DC event rates, respectively.}
\label{Baksan_10_kt}       
\end{figure}

The LBNT is the project of the scintillation telescope at the the Baksan Neutrino Observatory in Russia. With this telescope, it is planned to achieve the record efficiency indicators of the order of $95 - 100\%$ for a registration energy range of $0-100$ MeV  \cite{Bayarto}. Since the detector is at the design stage, the structure and physics of the backgrounds are still being discussed, and the effective volume of the detector is yet to be fixed. The detection technique of LBNT is similar to JUNO. The LBNT is expected to be protected from most backgrounds, except for NC and CC. It is estimated that the background of atmospheric neutrinos through NC and CC will be at the level of no more than $\sim 1$ and $\sim 2$ events per kiloton per year in the range up to 100 MeV, respectively. NC backgrounds dominate in energy range of 10-30 MeV (no more than 0.2 events per kiloton per year)  \cite{Bayarto}. These estimates included the possibility of using the signature of events associated with these backgrounds and analyzing the pulse shape to suppress the background. Accordingly, 10 kt for the fiducial volume and $95\%$ for the  efficiency were used for the calculations. Estimated event rate is shown in Fig. \ref{Baksan_10_kt}.
\section{Summary and Discussions}
\begin{figure}
\includegraphics[scale=0.51]{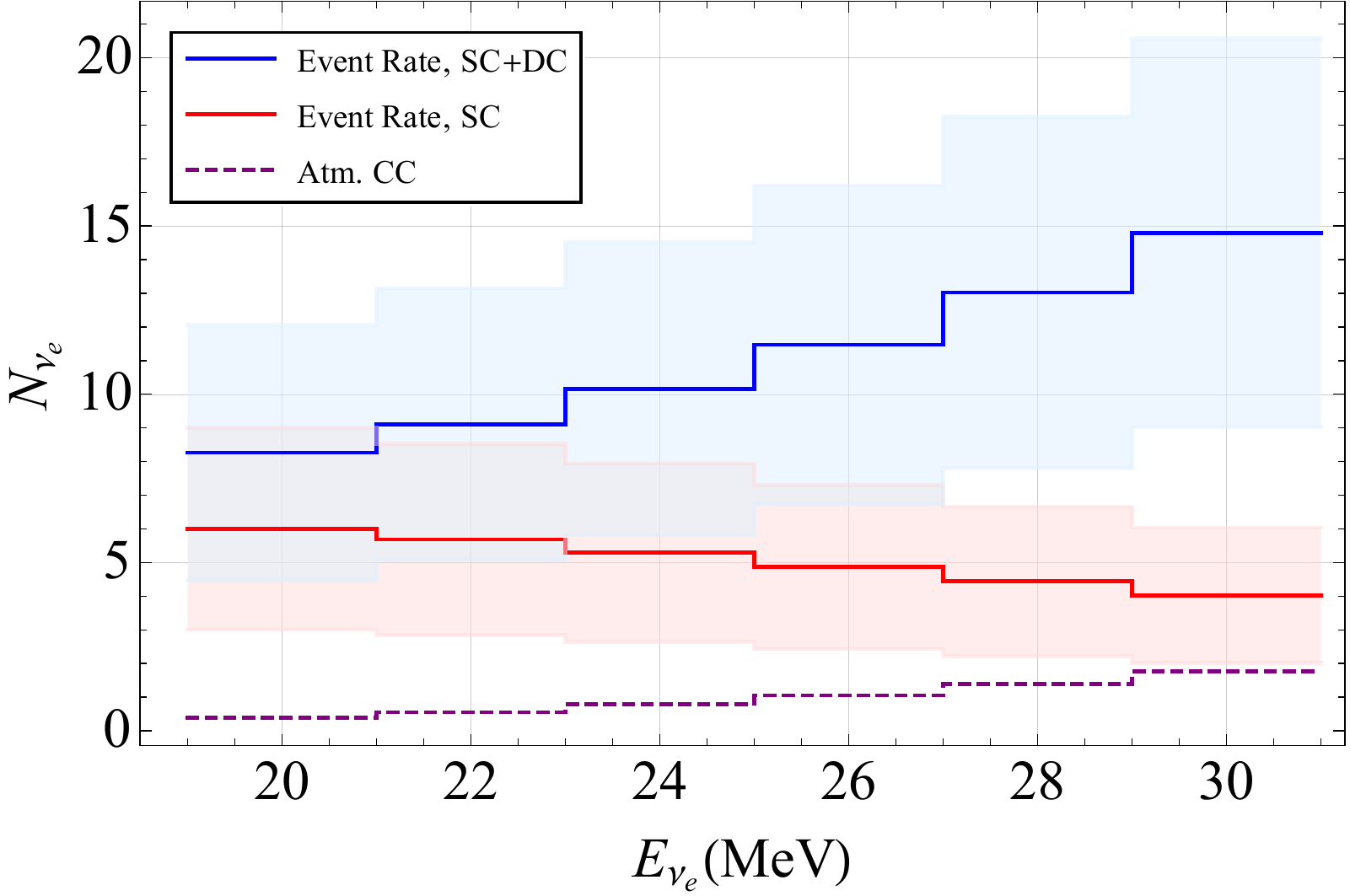}
\caption{Estimated number of events in 20 years for DUNE as a function of neutrino energy. Red and blue lines are SC and SC+DC with $\lambda = 10\%$ event rates, respectively. The dashed purple line is the atmospheric CC background.}
\label{DUNE_10}      
\end{figure}
\begin{figure}
\includegraphics[scale=0.51]{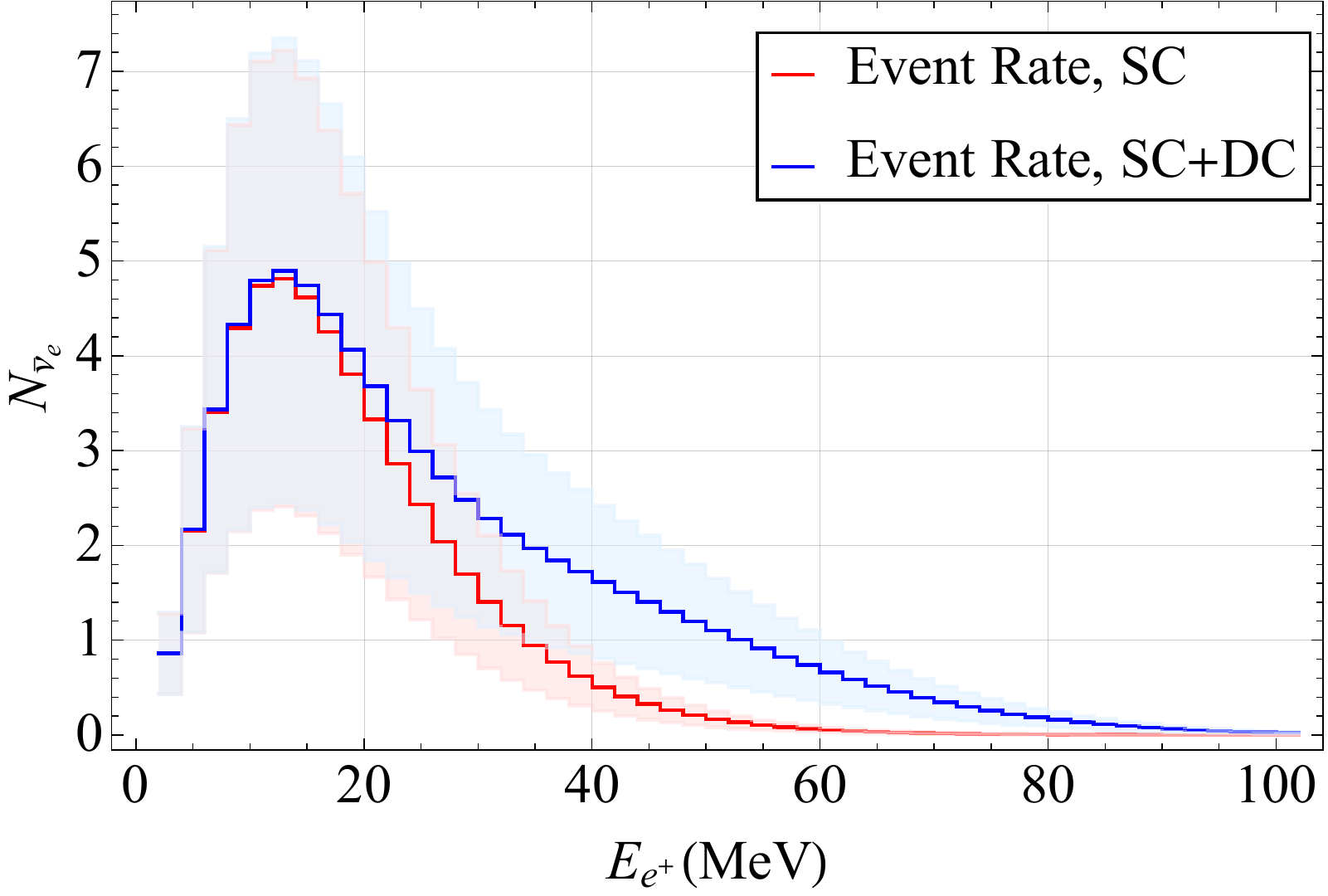}
\caption{Estimated number of events in 20 years for the LBNT as a function of positron energy. Red and blue lines are SC and SC+DC with $\lambda = 15\%$ event rates, respectively.}
\label{Baksan_15}       
\end{figure}
The details for calculating the DSNB are not fully understood yet. In particular, the CCSNe rate, its normalization and the IMF should be clarified in the future. It should be noted that only generic double and standard collapse models have been considered in this paper. There are alternative models, for instance, the magneto-rotational collapse mechanism \cite{Bisnovatyi_Kogan_2018}, models that take into account the formation of black holes during the collapse of massive stars \cite{Tabrizi_2021}, \cite {2003}, \cite {2012} which can be analyzed in a similar way. By analyzing the spectrum for future detectors, it is possible to determine the best strategy to determine the true collapse mechanism or to obtain constraints on scenarios.

HK is a huge water Cherenkov neutrino detector under construction \cite{protocollaboration2018hyperkamiokande}. It is supposed to work at full capacity within the upcoming decade, earlier than all other detectors considered here. Nevertheless, this detector unfortunately is unable to distinguish the DC model from the SC anywhere in its working energy range (Fig. \ref{HKgraph}).

JUNO is a neutrino experiment still at the project stage. The backgrounds in the energy ranges from  $\sim 30$ MeV up to $\sim 100$ MeV are not published yet, but, nevertheless, JUNO is a suitable detector for our purposes in terms of background reduction against backgrounds which gives accuracy and allows us to distinguish the DC from the background. According to the available data (Fig. \ref{JUNOgraph}), we can conclude that the main differences between the models are in energy range higher than $\sim 30$ MeV.

DUNE is very different from other detectors discussed in 
this work. Since it uses the argon reaction 
(\ref{ArgonReaction}) as a detection channel, so DUNE is 
sensitive to electron neutrinos especially. Fig. \ref{Fluxes} shows that the main differences between the DC and the SC should be in the sector of electron neutrinos (it is easy to see that in the scenario of the DC the number of 
electron neutrinos is much larger). In this work, we used a naive approximation of the cross-section for the reaction with argon which leads to additional inaccuracies. The DUNE is also at the design stage and there is 
not so much information on it today. Nevertheless, the 
differences in the number of events by an order of magnitude 
in the case of the DC allows us to conclude that DUNE will be very important for 
distinguishing between models. 

The main advantage of the LBNT is the background supression. This 
telescope is located far from nuclear power plants which makes it possible to register low-energy neutrinos. Also, this telescope is supposed to be built deep under the protection of a mountain shield. These factors and the experimental possibilities, which were described above, make it possible to almost get rid of backgrounds. Moreover, this detector is expected to achieve high accuracy. It should be noted that linear alkylbenzene is a relatively cheap filler which leads to an 
optimistic conclusion on the implementation of this project. However, this detector is at early design stage which introduces a lot of inaccuracies with its parameters. At least, it is not yet clear what the detector volume and efficiency will be. However, due to the very large registered energy range and small backgrounds (Fig. \ref{Baksan_10_kt}), one can confidently distinguish between the DC and the SC in the range of $30 - 60$ MeV, even with the help of electron antineutrino detection.

It is also reasonable to assume that the collapse model may depend on a particular star. As noted in the work \cite{Postnov_2016}, the number of supernovae collapsing in a double collapse maybe small. Simplified, if only the SC and the DC are found in nature, then the event rate is the sum of the event rates from the SC and the DC

\begin{equation}
    N_{DSNB} = \lambda \cdot N_{DC} + (1-\lambda) \cdot N_{SC}, 
    \label{final}
\end{equation}
where $N_{DC}$ and $N_{SC}$ are the event rates of the DC and the SC, respectively, $\lambda \in [ 0, 1]$.

Although the event rates in DUNE (Fig. \ref{DUNEDC}, \ref{DUNESM}) and the LBNT (Fig. \ref{Baksan_10_kt}), if $ \lambda = 1  (\mbox{or }0)$, give us a comprehension of the nature of the collapse according to the DC (SC) scenario, but if $\lambda$ is not equal to 1 (or 0), then the event rates will form a linear combination.

The combined model (\ref{final}) can be distinguished from the SC within the detectors capabilities only if $\lambda \gtrsim 0.1$. The only detectors that are able to distinguish between these two models are DUNE and the LBNT. Namely, DUNE is able to conclusively distinguish between the models at the energies of $E \gtrsim 30$ MeV and the DC contribution of $\lambda \gtrsim 0.1$ (Fig. \ref{DUNE_10}), and the LBNT distinguishes the models effectively at  $E \gtrsim 50$ MeV and $\lambda \gtrsim 0.15$ (Fig. \ref{Baksan_15}).

Despite all the discussed limitations, the present study demonstrates that the SC+DC model can be tested by observations of DSNB within upcoming decades. The choice of the strategy to test it should use the fact that DC models predict higher neutrino energies and a substantial $\nu_e$ flux, so the combination of DUNE ($\nu_e$) and LBNT (high energies) data is the most promising approach.

\begin{acknowledgments}
The Authors are grateful to Sergey Troitsky for the proposing the original idea and for fruitful discussions on the manu\-script. The Authors thank Bayarto Lubsandorzhiev for helpful remarks on the Large Baksan Neutrino Telescope, Maxim Libanov and Yury Kudenko for careful reading of the text and useful remarks. This work was supported by the Ministry of Science and Higher Education of the Russian Federation under the contract 075-15-2020-778 (state project “Science”).
\end{acknowledgments}
\bibliography{dsnb}
\end{document}